\documentclass[12pt,a4paper]{article}
\usepackage{graphicx}
\usepackage{amsthm}
\theoremstyle{plain}

\title{The Simple Equations Method (SEsM) and the use of  exponential functions for obtaining simple and multisoliton solutions of some nonlinear partial differential equations}
\author{Nikolay K. Vitanov}
\date{Institute of Mechanics, Bulgarian Academy of Scienecs, Acad. G. Bonchev Str., Bl. 4, 1113 Sofia, Bulgaria}
\begin{document}
\maketitle

\begin{abstract} 
We discuss the last version as well as applications of a  method for obtaining exact solutions of nonlinear partial differential equations. As this version is based on more than one simple equation 
we call it Simple Equations Method (SEsM). SEsM
contains as particular case the Modified Method of Simplest Equation (MMSE) for the
case when we use one simple equation and the solution is searched as power
series of the solution of the simple equation. SEsM contains as particular cases many
other methodologies for obtaining exact solutions of nonlinear partial differential equations.
We demonstrate that SEsM can lead to multisoliton solutions of integrable nonlinear partial differential equations and in addition we demonstrate that  SEsM keeps the property of the Modified Method of Simplest Equation to lead to exact solutions of nonitegrable  nonlinear partial differential equations.  
\end{abstract}
\section{Introduction}
The evolution of scientific research happened in such direction that
in the course of the time researchers started to study systems possessing larger complexity \cite{a1}-
\cite{a4}. These systems have many interesting features and one of the most interesting  feature of
many complex systems is their nonlinearity \cite{n1} - \cite{n8}. The nonlinearity of the complex systems can be studied, e.g.,  by means of time series analysis or by means of models based on differential equations \cite{t1} - \cite{t10} . In many cases the corresponding model equations are nonlinear partial differential equations.
The research on  the methodology for obtaining exact solutions of nonlinear partial differential
equations began by study of transformations that  transform the solved nonlinear partial differential equation to a linear differential equation. One example is  the Hopf-Cole transformation \cite{hopf}, \cite{cole}  which transforms the nonlinear Burgers equation to the linear heat equation.   The  attempts for obtaining similar transformations led to the development of the methodology of the 
\emph{Method of Inverse Scattering Transform} \cite{ablowitz} - \cite{gardner}. 
In addition Hirota  developed a direct method for obtaining of  exact solutions of NPDEs -  \emph{Hirota method} \cite{hirota}, \cite{hirota1}. This method 
is based on bilinearization of the solved nonlinear partial differential equation by means of appropriate transformations. Truncated Painleve expansions  lead to many such transformations \cite{tabor} - \cite{wtk}.  Kudryashov \cite{k3} studied the possibility
for obtaining exact solutions of NPDEs by a truncated Painleve expansion where the truncation happens after the "constant term".  Kudryashov  formulated   the \emph{Method of Simplest Equation (MSE)} \cite{k05} based 
on determination of singularity order $n$ of the solved NPDE and searching of
particular solution of this equation as series containing powers of solutions
of a simpler equation called \emph{simplest equation}. The methodology was extended \cite{kl08}
and applied  for obtaining traveling wave solutions of  nonlinear partial differential equations
(see, e.g., \cite{k5a} - \cite{k12}). 
\par
Our way to the methodology of SEsM was long. Some elements of the methodology can be observed in
our articles written 25 - 30 years ago \cite{mv1} - \cite{mv5}. What followed started in 2009
\cite{1}, \cite{2} and continued in 2010 by the use of the ordinary differential 
equation of Bernoulli as simplest equation \cite{v10} and by application of the methodology of the
Modified Method of Simplest Equation to ecology
and population dynamics \cite{vd10}. In these publications we have used the concept of the balance equation. 
Note that the version  called \emph{Modified Method of Simplest Equation - MMSE} \cite{vdk}, \cite{v11} based on determination of the kind
of the simplest equation and truncation of the series of solutions of the simplest equation by means 
of application of a balance equation is equivalent of the \emph{Method of Simplest Equation}. 
Up to now our contributions to the methodology and its application have been connected to the \emph{MMSE} 
\cite{v11a} - \cite{vdv17}. We note especially the article \cite{vdv15} where we have extended the methodology 
of the \emph{MMSE} to simplest equations of the class
\begin{equation}\label{sf}
\left (\frac{d^k g}{d\xi^k} \right)^l = \sum \limits_{j=0}^{m} d_j g^j
\end{equation}
where $k=1,\dots$, $l =1,\dots$, and $m$ and $d_j$ are parameters. The solution of Eq.(\ref{sf}) defines
a special function that contains as particular cases, e.g.,: (i) trigonometric functions; (ii) hyperbolic functions; (iii) elliptic functions of Jacobi; (iv) elliptic function of Weierstrass.
\par 
In the course of the time we started to extend the methodology of the Modified Method of Simplest Equation.
The last version of the methodology 
is connected to the possibility of use of more than one simplest equation. This modification may be called  
MMSEn (Modified Method of Simplest Equation based on $n$ simplest equations)
but it is better to call it SEsM - Simple Equations Method. The reason for this is that the used simple equations are more simple than the solved nonlinear partial differential equation but these simple equations in fact can be quite complicated.   
A variant of SEsM based on two simple equations was applied in \cite{vd18}
and the first description of the methodology was made in \cite{v19} and then
in \cite{v19a}. Mor more applications of particular cases of the methodology see
\cite{n17} - \cite{iv19}. 
\par
The text below is organized as follows. We discuss the SEsM  in Sect 2 and 
show that the SEsM is capable to lead to multisoliton solutions of
nonlinear partial differential equations. The example will be based on the
Korteweg - de Vries equation. In Sect.3 we demonstrate that SEsM keeps
the property of MMSE to lead to exact solutions of nonitegrable nonlinear
partial differential equations. Several concluding remarks are summarized in Sect. 4.
We note that all the solutions below are based on exponential functions and thus
we emphasize the importance of the exponential function for the efforts for obtaining
exact solutions of nonlinear partial differential equations.
\section{Description of the Simple equation (SEsM) with application for obtaining multisoliton solutions}
The schema of SEsM is as follows. We have to solve a nonlinear partial differential equation 
\begin{equation}\label{eqx}
{\cal{D}}u(,\dots)=0
\end{equation}
where ${\cal{D}}(u,\dots)$ is a relationship containing the function $u(x,...,t)$
and some of its derivatives   ($u$ can be a function of more than 1 spatial coordinate).
The methodology of SEsM has the following 7 steps.
\begin{description}
	\item[1.)]
	We apply a transformation
	\begin{equation}\label{m1}
	u(x,\dots,t)=Tr(F(x,\dots,t))
	\end{equation}
	where $Tr(F)$ is a function of another function  $F$. In general
	$F(x,\dots,t)$ is a function of the spatial variables as well as of the time. 
	The transfomation has to transform the nonlinearity of the solved differential equation to
	more treatable kind of nonlinearity or the transformation may even remove the nonlinearity.
	Several examples for  the form of the transformation $Tr(F)$ are
	\begin{itemize}
	\item the Painleve expansion \cite{hirota}, \cite{k3}, 
	\item $u(x,t)=4 \tan^{-1}[F(x,t)]$ for the case of the 
	sine - Gordon equation ,
	\item  $u(x,t) = 4 \tanh^{-1}[F(x,t)]$ for the case of sh-Gordon (Poisson-Boltzmann 
	equation) (for applications of the last two transformations see, e.g. \cite{mv1} - \cite{mv5}),
	\item another transformation.
	\end{itemize}
	In many particular cases one may skip this step (then we have just $u(x,\dots,t)=F(x,\dots,t)$) 
	but in numerous cases the step is necessary
	for obtaining a solution of the studied nonlinear PDE. The application of Eq.(\ref{m1}) to 
	Eq.(\ref{eqx}) leads to a nonlinear PDE for the function $F(x,\dots,t)$. We note that no general form
	of the transformation $Tr(F)$ is known.
	\item[2.)]
	The function $F(x,\dots,t)$ is written as a function of other functions $f_1,\dots,f_N$. These
	functions are  connected to solutions of some differential equations (these equations can be partial 
	or ordinary differential equations) that are more simple than Eq.(\ref{eqx}).  
	The possible values of $N$ are $N=1,2,\dots,\infty$.
	The form of the function $F(f_1,\dots,f_N)$ is not prescribed and can be given by different
	relationships, e.g.,
	\begin{itemize} 
		\item
	\begin{eqnarray}\label{m2}
	F &=& \alpha + \sum \limits_{i_1=1}^N \beta_{i_1} f_{i_1} + \sum \limits_{i_1=1}^N  \sum \limits_{i_2=1}^N 
	\gamma_{i_1,i_2} f_{i_1} f_{i_2} + \dots + \nonumber \\
	&&\sum \limits_{i_1=1}^N \dots \sum \limits_{i_N=1}^N \sigma_{i_1,\dots,i_N} f_{i_1} \dots f_{i_N}
	\end{eqnarray}
	where $\alpha,\beta_{i_1}, \gamma_{i_1,i_2}, \sigma_{i_1,\dots,i_N}\dots  $ are parameters.
	\item
	 or $F(f_1,\dots,f_N)$ can have another form.
	\end{itemize}
	We shall use the form given by Eq.(\ref{m2}) below. Note that the relationship (\ref{m2}) contains as particular case
	the relationship used by Hirota \cite{hirota}. In addition 
	the power series $\sum \limits_{i=0}^N \mu_n f^n$ (where
	$\mu$ is a parameter) used in the previous versions of the methodology based on 1 simple equation (and called  Modified Method of Simplest Equation) are a particular case of the relationship (\ref{m2}) too.
	\item[3.)] 
	In general each of the functions $f_1,\dots,f_N$ is a solution of a partial differential equation. These equations are more simple than the
	solved nonlinear partial differential equation. There are two
	possibilities: one may use solutions of the simple partial differential equations if such solutions are available, or one may transform the more simple partial differential equations 
	by means of appropriate ans{\"a}tze (e.g.,  traveling-wave ans{\"a}tze such as 
	$\xi = \hat{\alpha} x + \hat{\beta} t$;  
	$\zeta = \hat{\mu} y + \hat{\nu}t \dots$). Then 
	the solved   differential equations for $f_1,\dots,f_N$ may be reduced to   differential equations 
	$E_l$, containing derivatives of one or several functions
	\begin{equation}\label{i1}
	E_l \left[ a(\xi), a_{\xi},a_{\xi \xi},\dots, b(\zeta), b_\zeta, b_{\zeta \zeta}, \dots \right] = 0; \ \
	l=1,\dots,N
	\end{equation}
	In many cases (e.g, if the equations for the functions $f_1,\dots$ are ordinary differential equations) one may skip this step 
	but the step may be necessary if the equations for $f_1,\dots$ are complicated partial differential equations.
	\item[4.)]
	We assume that	the functions $a(\xi)$, $b(\zeta)$, etc.,  are  functions of 
	other functions, e.g., $v(\xi)$, $w(\zeta)$, etc., i.e.
	\begin{equation}\label{i1x}
	a(\xi) = A[v(\xi)]; \ \ b(\zeta) = B[w(\zeta)]; \dots
	\end{equation} 
	The kinds of the functions $A$ , $B$, $\dots$ are not prescribed. 
	Often one uses a finite-series relationship, e.g., 
	\begin{equation}\label{i2}
	a(\xi) = \sum_{\mu_1=-\nu_1}^{\nu_2} q_{\mu_1} [v (\xi)]^{\mu_1}; \ \ \ 
	b(\zeta) = \sum_{\mu_2=-\nu_3}^{\nu_4} r_{\mu_2} [w (\zeta)]^{\mu_2}, \dots 
	\end{equation}
	where $q_{\mu_1}$, $r_{\mu_2}$, $\dots$ are parameters.
	However other kinds of relationships may be used too. 
	\item[5.)]
	The functions  $v(\xi)$, $w(\zeta)$, $\dots$ 
	are solutions of simple ordinary differential equations.
	For several years we have used particular case of the described 
	methodology that was based on use of one simple equation. This simple equation was called simplest equation and the
	methodology based on one equation was called Modified Method of Simplest Equation. SEsM contains the Modified Method of Simplest Equation as  particular cease (as one of the numerous particular cases of the SEsM methodology).
	\item[6.)]
	The application of the steps 1.) - 5.) to Eq.(\ref{eqx}) transforms its  left-hand side. We
	consider the case when the result of this transformation  is a function that 
	is a sum of terms where each 
	term contains some function multiplied by a coefficient. This coefficient contains some of the 
	parameters of the solved equation and some of the parameters of the solution. In the most cases
	a balance procedure must be applied in order to ensure that the above-mentioned relationships
	for the coefficients contain more than one term (e.g., if the result of the transformation 
	is a polynomial then the balance procedure has to ensure that the coefficient of each 
	term of the polynomial is a relationship that contains at least two terms).
	This balance procedure may lead to one or more additional relationships among the parameters 
	of the solved equation and parameters of the solution. These relationships are called 
	\emph{balance equations}. 
	\item[7.)]
	We may obtain a nontrivial solution of Eq. (\ref{eqx})  if all coefficients mentioned in Step 6.) are
	set to $0$. This condition usually leads to a system of nonlinear algebraic equations for the 
	coefficients of the solved nonlinear PDE and for the coefficients of the solution. Any nontrivial 
	solution of this algebraic system leads to a solution the studied  nonlinear partial differential 
	equation. Usually the above system of algebraic equations contains many equations that have to 
	be solved with the help of   a computer algebra system. Sometimes the system is quite complicated and
	even the computer algebra systems are unable to solve it.
\end{description}
Let us now show several examples of the application of the methodology.
\subsection{The multisoliton solution of the Korteweg - de Vries equation}
Let us show that the SEsM methodology allows obtaining of the multi-soliton solution of the
	Korteweg-de Vries equation for the particular case when $N$ simple equations
	for exponential functions are used and the solution is constructed by use
	of the particular case of the function (\ref{m2}).  In order to do this we
consider the following version of the Korteweg - de Vries equation
	\begin{equation}\label{a1}
	u_t + \sigma u u_x + u_{xxx}=0
	\end{equation}
	where $\sigma$ is a parameter.
	We consider the particular (classical) case $\sigma=-6$ and use the transformantion
	\begin{equation}\label{tr1}
	u(x,t) = -2 \frac{\partial^2}{\partial x^2} \ln F(x,t)
	\end{equation}
	Let us consider the particular case of Eq.(\ref{m2}) where $\alpha=1$, $\beta_i=1$, $\gamma_{ij} \ne 0$ for $i \ne j$, $\delta_{ijk}\ne 0$ for $i \ne j \ne k$,$\dots$. We obtain
	\begin{eqnarray}\label{x2}
	F &=& 1 + \sum \limits_{i=1}^N  f_{i} + \sum \limits_{(i \ne j),i,j=1}^N   
	\gamma_{i,j} f_{i} f_{j} +  
	\sum \limits_{(i \ne j \ne k),i,j,k=1}^N   
	\delta_{i,j,k} f_{i} f_{j} f_{k} +\dots + \nonumber \\
	&& \sigma_{1,\dots,N} f_{1} \dots f_{N}
	\end{eqnarray}
	Let us consider $n$ simple equations of the kind
	\begin{equation}\label{x3}
	\frac{df_i}{d \xi_i} = f_i,  \ \ \ \xi_i = p_i x - \omega_i t - \xi_{i0}
	\end{equation}
	which solution is $f_i(\xi_i) = \exp(\xi_i)$. We substitute Eqs.(\ref{x3}) in Eq.(\ref{x2}) and select
	the coefficients $\gamma, \delta, \dots, $ in appropriate way in order to  obtain the relationship
	\begin{equation}\label{x4}
	F= 1+ \sum \limits_{n=1}^N \sum \limits_{{N} \choose{n}} \Bigg[\exp(\xi_{i_1}+ \dots+\xi_{i_N}) 
	\prod \limits_{k<l}^{(n)} \frac{(p_{ik}-p_{il})^2}{(p_{ik}+p_{il})^2}\Bigg]
	\end{equation}
	Above $\sum \limits_{{N} \choose{n}}$ is the sum over all combinations of $n$ elements taken from the
	set of $N$ elements and $\prod \limits_{k<l}^{(n)}$ is the product of all possible combinations of the n elements with the  condition $k <l$. The substitution of 
	\begin{equation}\label{x5}
	u(x,t) = -2 \frac{\partial^2}{\partial x^2} \ln \Bigg \{ 1+ \sum \limits_{n=1}^N \sum \limits_{{N} \choose{n}} \Bigg[\exp(\xi_{i_1}+ \dots+\xi_{i_N}) 
	\prod \limits_{k<l}^{(n)} \frac{(p_{ik}-p_{il})^2}{(p_{ik}+p_{il})^2}\Bigg] \Bigg \}
	\end{equation}
	in Eq.(\ref{a1}) ($\sigma = -6$) reduces this equation to the system of algebraic relationships
	\begin{eqnarray}\label{x6}
	\sum \limits_{l=0}^n \sum \limits_{{l} \choose{n}}\Bigg[\prod_{k<m}^{(l)} \frac{(p_{ik}-p_{im})^2}{(p_{ik}+p_{im})^2}\Bigg]
	\Bigg[\prod \limits_{k>l,k<m}^{(n)} \frac{(p_{ik}-p_{im})^2}{(p_{ik}+p_{im})^2}\Bigg] \Bigg[(-p_{i_1}-\dots-p_{i_l}+p_{i_{(l+1)}}+ \dots+ \nonumber \\
	p_{i_n})(-p_{i_1}-\dots-p_{i_l}+p_{i_{(l+1)}}+ \dots+ 
	p_{i_n})^3 -(-p_{i_1}^3-\dots-p_{i_l}^3+p_{i_{(l+1)}}^3+ \dots+ \nonumber \\
	p_{i_n}^3) \Bigg] =0
	\nonumber \\
	\omega_i = p_i^3\nonumber \\
	\end{eqnarray}
	for $n=1,\dots,N$. The nontrivial solution of this algebraic system leads to solitary or
	multisoliton solution of the Korteweg-de Vries equation.
\par
In order to make the example more clear we shall obtain
the two-soliton solution of the Korteweg - de Vries equation. 
Let us consider Eq. (\ref{a1}) and let $\sigma$  be a parameter.
 The 7 steps of the SEsM  are as follows
\begin{description}
	\item[1.)] \emph{The transformation}\\
	We set $u=p_x$ in Eq.(\ref{a1}). The result is integrated  and we apply the transformation
	$p=\frac{12}{\sigma} (\ln F)_x$. The result is
	\begin{equation}\label{a2}
	FF_{tx} + FF_{xxxx} - F_t F_x  + 3 F_{xx}^2 - 4 F_x F_{xxx} =0
	\end{equation}
	\item[2.)] \emph{Relationship among $F(x,t)$ and two functions $f_{1,2}$ that will be connected below to two simplest equations}\\ 
	We shall use two functions
	$f_1(x,t)$ and $f_2(x,t)$ and the relationship for $F$ is assumed to be a particular case of 
	Eq.(\ref{m2}) namely
	\begin{equation}\label{a3}
	F(x,t) = 1 + f_1(x,t) + f_2(x,t) + c f_1(x,t) f_2(x,t) 
	\end{equation}
	where $c$ is a parameter. We note again that Eq.(\ref{m2}) contains as particular case the
	relationship used by Hirota in \cite{hirota}. The substitution of Eq.(\ref{a3}) in Eq.(\ref{a2}) leads to
	\begin{eqnarray}\label{a4}
	&& f_{1xxxx}+f_{2xxxx}+3f_{1xx}^2+cf_{1xt} f_2 + f_{1xt} - f_{1t} f_{1x}-f_{1t} f_{2x} -
	\nonumber \\
	&&4 f_{1x} f_{1xxx} -  f_{1x} f_{2t} - 4 f_{1x} f_{2xxx} + 6 f_{1xx} f_{2xx} - 4 f_{1xxx} f_{2x} - f_{2t} f_{2x} - \nonumber \\
	&& 4 f_{2x} f_{2xxx} +  f_2 f_{1xt} + f_2 f_{1xxxx} + f_2 f_{2xt} + f_2 f_{2xxxx} + f_1 f_{1xt} + 
	\nonumber \\
	&& f_1 f_{1xxxx} + f_1 f_{2xt} +
	f_1 f_{2xxxx} + f_{2xt} + c^2 f_2^2 f_1 f_{1xt} + c^2 f_2^2 f_1 f_{1xxxx} - \nonumber \\
	&& c^2 f_2^2  f_{1t} f_{1x} - 4f_2^2 c^2 f_{1x} f_{1xxx} + c^2 f_2 f_1^2 f_{2xt} + c^2 f_2 f_1^2 f_{2xxxx} - \nonumber \\
	&& 12 c^2 f_2 f_{1x}^2 f_{2xx} - c^2 f_1^2 f_{2t} f_{2x} - 
	4 c^2 f_1^2 f_{2x} f_{2xxx} - 12 c^2 f_1 f_{1xx} f_{2x}^2 + \nonumber \\
	&& 2 c f_2 f_1  f_{1xt} + 2 c f_2 f_1  f_{1xxxx} + 2 c f_2 f_1  f_{2xt} + 2 c f_2 f_1  f_{2xxxx} - 2 c f_2  f_{1t} f_{1x} - \nonumber \\
	&& 8 c f_2  f_{1x} f_{1xxx} + 12 c f_2  f_{1xx} f_{2xx} + 12 c f_1  f_{1xx} f_{2xx} - 2 c f_1  f_{2t} f_{2x} - \nonumber \\
	&& 8 c f_1  f_{2x} f_{2xxx}+ 12 c^2 f_{1x}^2 f_{2x}^2 - 12 c f_{1x}^2 f_{2xx} - 12 c f_{1xx} f_{2x}^2 + 3 c^2 f_2^2 f_{1xx}^2 + \nonumber \\
	&& 3 c^2 f_1^2 f_{2xx}^2 + c f_2^2  f_{1xt} + c f_2^2  f_{1xxxx} + 6 c f_2 f_{1xx}^2 + c f_1^2 f_{2xt} + \nonumber \\
	&& c f_1^2 f_{2xxxx} + 6 c f_1 f_{2xx}^2 + c f_1 f_{2xt} + c f_1 f_{2xxxx} + 6 c f_{1xx} f_{2xx} + 4 c f_{1x} f_{2xxx}+ \nonumber \\
	&& c f_{1x} f_{2t} + c f_{1xxxx} f_2 + 12 c^2 f_2 f_1 f_{1xx} f_{2xx} + c f_{1t} f_{2x} + 4 c f_{1xxx} f_{2x} + 3 f_{2xx}^2 =0
	\nonumber \\
	\end{eqnarray}
	\item[3.)] \emph{Equations for the functions $f_1(x,t)$ and $f_2(x,t)$}\\
	The structure of Eq.(\ref{a4}) allow us to assume  a very simple form of the equations for the functions $f_{1,2}$:
	\begin{eqnarray}\label{a5}
	\frac{\partial f_1}{\partial x} &=& \alpha_1 f_1; \ \ \ \frac{\partial f_1}{\partial t} = \beta_1 f_1; \nonumber \\
	\frac{\partial f_2}{\partial x} &=& \alpha_2 f_2; \ \ \ \frac{\partial f_2}{\partial t} = \beta_2 f_2;
	\end{eqnarray}
	This choice will transform Eq.(\ref{a4}) to a polynomial of $f_1$ and $f_2$. Further we assume that
	$\xi = \alpha_1 x + \beta_1 t + \gamma_1$ and $\zeta = \alpha_2 x + \beta_2 t + \gamma_2$ and
	\begin{equation}\label{a6}
	f_1(x,t) = a(\xi); \ \ \ f_2(x,t) = b(\zeta)
	\end{equation}
	Above $\alpha_{1,2}$, $\beta_{1,2}$ and $\gamma_{1,2}$ are parameters.
	\item[4.)] \emph{Relationships connecting  $a(\xi)$ and $b(\zeta)$ to the functions $v(\xi)$ and
		$w(\zeta)$ that are solutions of the simplest equations} \\
	In the discussed here case the relationships are quite simple. We can use Eq.(\ref{i2}) for the cases
	$\mu_1 = \nu_2 = 1$ and $\mu_2 = \nu_4 = 1$. The result is
	\begin{equation}\label{a7}
	a(\xi) = q_1 v(\xi); \ \ \ b(\zeta) = r_1 w(\zeta)
	\end{equation}
	\item[5.)] \emph{Simple equations for $v(\xi)$ and $w(\zeta)$}\\
	The simple equations are
	\begin{equation}\label{a8}
	\frac{dv}{d\xi} = v; \ \ \ \frac{dw}{d\zeta} = w
	\end{equation}
	and the corresponding solutions are
	\begin{equation}\label{a9}
	v(\xi) = \omega_1 \exp (\xi); \ \ \ w(\zeta)  = \omega_2 \exp(\zeta)
	\end{equation}
	Below we shall assume that the parameters $\omega_{1,2}$ are  included in the parameters
	$q_1$ and $r_1$ respectively. We shall assume also that $q_1$ and $r_1$ can be included in $\xi$ and $\zeta$.
	\item[6.)] \emph{Transformation of Eq.(\ref{a4})}\\
	Let us substitute Eqs.(\ref{a5}) - (\ref{a8}) in Eq.(\ref{a4}). The result is a sum of exponential functions
	and each exponential function is multiplied by a coefficient. Each of these coefficients is a relationship
	containing the parameters of the solution and all of the relationships contain more than one term. Thus
	we don't need to perform a balance procedure.
	\item[7.)] \emph{Obtaining and solving the system of algebraic equations}\\
	The system of algebraic equations is obtained by setting of above-mentioned relationships to $0$.
	Thus we obtain the following system:
	\begin{eqnarray}\label{a10}
	&& \alpha_1^3 + \beta_1 = 0, \nonumber \\
	&& \alpha_2^3 + \beta_2 = 0, \nonumber \\
	&& (c+1) \alpha_1^4 + 4 \alpha_2 (c-1) \alpha_1^3 + 6 \alpha_2^2 (c+1) \alpha_1^2 + [(4c-4)\alpha_2^3 + (\beta_1 + \beta_2) c + \nonumber \\
	&& \beta_1 - \beta_2] \alpha_1 + [(c+1)\alpha_2^3 + (\beta_1 + \beta_2) c - \beta_1 + \beta_2] \alpha_2 = 0.
	\end{eqnarray}
	The non-trivial solution of this system is 
	\begin{eqnarray}\label{a11}
	\beta_1 = -\alpha_1^3; \ \ \beta_2 = -\alpha_2^3; \ \
	c = \frac{(\alpha_1 - \alpha_2)^2}{(\alpha_1 + \alpha_2)^2}
	\end{eqnarray}
	and the corresponding solution of Eq.(\ref{a1}) is
	\begin{eqnarray}\label{a12}
	u(x,t) &=& \frac{12}{\sigma} \frac{\partial^2}{\partial x^2} \Bigg[ 1+ \exp \Big(\alpha_1 x - \alpha_1^3 t + \gamma_1 \Big) + \exp \Big(\alpha_2 x - \alpha_2^3 t + \gamma_2 \Big) + \nonumber \\
	&& \frac{(\alpha_1 - \alpha_2)^2}{(\alpha_1 + \alpha_2)^2}
	\exp \Big( (\alpha_1 + \alpha_2)x - (\alpha_1^3 + \alpha_2^3)t + \gamma_1 + \gamma_2 \Big) \Bigg]
	\end{eqnarray}
	Eq.(\ref{a12}) describes the bisoliton solution of the Korteweg - de Vries equation.
\end{description}
\par 
The use of more simple equations for exponential functions will lead to  solutions, containing more solitons.
Thus we have proven that the methodology of SEsM  is capable to search for
complicated solitary wave solutions solutions of nonlinear PDEs. This capability is acquired on the 
basis of the 
possibility of use of more than one simple equation. The relationship (\ref{m2}) 
can be used also for obtaining  exact solution of nonintegrable nonlinear PDEs. This will be 
demonstrated in the following section. 
\section{Exponential functions and nonintegrable equations}
Let us consider the simple equation
\begin{equation}\label{e1}
\frac{dv}{d \xi} = v^2 -v
\end{equation}
It has the solution
\begin{equation}\label{e2}
v(\xi) = \frac{1}{1+\exp(\xi)}
\end{equation}
We can search solutions of the solved nonlinear partial differential equations ${\cal{D}}u(\xi)=0$
as
\begin{equation}\label{e3}
u(\xi) = \sum \limits_{i=0}^N a_i v^i
\end{equation}
This was proposed by Kudryashov in 2012 \cite{k12}. Now let us show that 
the above method of Kudryashov is particular case of the SEsM methodology for the case
of use of 1 simple equation and representation of the searched solution as a polynomial of the
solution of the simple equation.	In order to do this
we consider the SEsM methodology for the case of lack of transformation, one
simple equation ( Eq.(\ref{e1})) and the polynomial representation (\ref{e3}) of the
solution of the equation by the solution of the simple equation.
Then the SEsM methodology is reduced to the method of Kudryashov. Hence the
method of Kudryashov is particular case of the SEsM methodology.
\par
Let us now use this particular case  to obtain an exact solution for a  particular case of  the equation
\begin{equation}\label{b1}
u_t + \left( \sum \limits_{k=0}^l \alpha_k u^k \right) u_x + \beta u_{xxx} + \gamma u^m u_{xxxxx}=0
\end{equation}
We apply the methodology for the particular case $u(x,t)=F(x,t)$ of
transformation (\ref{m1}). The steps of the methodology are as follows
\begin{description}
	\item[1.)] \emph{The transformation}\\
	We shall  use a  particular case of transformation (\ref{m1}), i.e. $u(x,t)=F(x,t)$.
	\item[2.)] \emph{Relationship among $F(x,t)$ and the functions $f_k(x,t)$}\\
	The function F(x,t) will be searched as a function of another function $f(x,t)$ and the corresponding relationship 
	is particular case of the relationship (\ref{m2})
	\begin{equation}\label{c1}
	F(x,t) = \sum \limits_{i=0}^N \gamma_i f(x,t)^i
	\end{equation}
	where $\gamma_i$ are parameters.
	\item[3.)] \emph{Equation for the function $f(x,t)$}\\
	The function $f(x,t)$ will be assumed to be a traveling wave
	\begin{equation}\label{c2}
	f(x,t) = a(\xi); \ \ \ \xi = \mu x + \nu t
	\end{equation}
	where $\mu$ and $\nu$ are parameters.
	\item[4.)] \emph{Representation of the function $a(\xi)$ by a function that is solution of a simplest equation}\\
	We shall not express further the function $a(\xi)$ through another function $v(\xi)$ and
	instead of this we shall assume that $a(\xi)$ is a solution of a simple equation of the class
	(\ref{a1}). 
	\item[5.)] \emph{The simple equation}\\
	Below we shall use a particular case of the following simple equation:
	\begin{equation}\label{c3}
	\frac{da}{d\xi}  = \sum \limits_{j=0}^{p} d_j a^j,
	\end{equation}
	\item[6.)] \emph{Transformation of Eq.(\ref{b1})}\\
	The substitution of Eqs. (\ref{c1}) and (\ref{c3}) in Eq.(\ref{b1})
	leads to a polynomial of $a(\xi)$ that contains the following maximum powers of the terms of Eq.(\ref{b1}) : $N+p-1$; $N+3(p-1)$; $Nm+n+5(p-1)$; $Nl+N+p-1$. 
	In order to obtain the system of nonlinear algebraic
	equations we have to write balance equations for these powers, i.e. in this case we have to balance 
	the largest powers:  
	$Nm+n+5(p-1)$ and $Nl+N+p-1$. This leads us to the balance equation
	\begin{equation}\label{c5}
	N(l-m) = 4(p-1)
	\end{equation}
	We note that $l$, $m$, $p$, $N$ have to be integers or $0$. We have $p>1$ and $l>m$. Then from Eq.(\ref{c5})
	\begin{equation}\label{c6}
	N = 4 \frac{p-1}{l-m}
	\end{equation}
	which means that  the equations of the class Eq.(\ref{b1}) may have solutions of the kind
	\begin{equation}\label{c7}
	u(x,t) = \sum \limits_{i=0}^{4 \frac{p-1}{l-m}} \gamma_i a(\xi)^i,
	\end{equation}
	where $\xi = \mu x + \nu t$ and $a(\xi)$ is a solution of the simple equation
	\begin{equation}\label{c8}
	\frac{da}{d\xi} = d_0 + \dots + d_p a^p.
	\end{equation}
	We note that $\frac{p-1}{l-m}$ must be an integer. Let us set $m=1$. The smallest possible value of $l$ in this case is $l=2$. Then
the equation (\ref{b1}) is reduced to
\begin{equation}\label{b1x}
u_t + \left( \alpha_0 + \alpha_1 u + \alpha_2 u^2 \right) u_x + \beta u_{xxx} + \gamma u u_{xxxxx}=0
\end{equation}
Next we shall use Eq.(\ref{c3}) as simple equation. From Eq.(\ref{c6}) we obtain for the balance equation
$N = 4 (p-1)$
which means that  the equations of the class Eq.(\ref{b1x}) may have solutions of the kind
\begin{equation}\label{cy1}
u(x,t) = \sum \limits_{i=0}^{4 (p-1)} \gamma_i a(\xi)^i,
\end{equation}
where $\xi = \mu x + \nu t$ and $a(\xi)$ is a solution of the simple equation
\begin{equation}\label{cy2}
\frac{da}{d\xi} = d_0 + \dots + d_p a^p.
\end{equation}
We shall consider the case $p=2$ below. In this case the application of the steps of the
methodology leads to a system of 13 nonlinear algebraic equations for the parameters of the
solution and the parameters of the equation. As we want to use the simple equation (\ref{e1})
we have to set $d_0=0$, $d_1=-1$ and $d_2=1$ which simplifies much the system of algebraic equations.
One nontrivial solution of this system is
\begin{eqnarray}\label{as}
\gamma_0 &=& - \frac{1}{15} \frac{6^{1/2} 35^{3/4} \beta \gamma}{(\alpha_2^3 \beta^3 \gamma^2)^{1/4}}; 
\ \
\gamma_1 = 4 \frac{6^{1/2} 35^{3/4} \beta \gamma}{(\alpha_2^3 \beta^3 \gamma^2)^{1/4}}; \nonumber\\
\gamma_2 &=& - 28 \frac{6^{1/2} 35^{3/4} \beta \gamma}{(\alpha_2^3 \beta^3 \gamma^2)^{1/4}}; \ \
\gamma_3 = 48 \frac{6^{1/2} 35^{3/4} \beta \gamma}{(\alpha_2^3 \beta^3 \gamma^2)^{1/4}}; \nonumber \\
\gamma_4 &=& - 24 \frac{6^{1/2} 35^{3/4} \beta \gamma}{(\alpha_2^3 \beta^3 \gamma^2)^{1/4}}; \ \
\mu = \frac{1}{3}\frac{6^{1/2}35^{1/4}(\alpha_2^3 \beta^3 \gamma^2)^{1/4}}{\alpha_2 \beta} \nonumber \\
\nu &=& - \frac{1}{3}\frac{6^{1/2}35^{1/4}\alpha_0(\alpha_2^3 \beta^3 \gamma^2)^{1/4}}{\alpha_2 \beta}; \ \
\alpha_1 = \frac{1}{10} \frac{6^{1/2} 35^{3/4} \beta \gamma}{(\alpha_2^3 \beta^3 \gamma^2)^{1/4}}; 
\end{eqnarray}
and the corresponding solution of the solved equation (\ref{b1x}) is
\begin{eqnarray}\label{fs}
u(\xi) &=& \frac{6^{1/2} 35^{3/4} \beta \gamma}{(\alpha_2^3 \beta^3 \gamma^2)^{1/4}} \Bigg \{ 
- \frac{1}{15} + 4 \Bigg[\frac{1}{1+\exp \Bigg[ \Bigg( \frac{1}{3}\frac{6^{1/2}35^{1/4}(\alpha_2^3 \beta^3 \gamma^2)^{1/4}}{\alpha_2 \beta} \Bigg)(x-\alpha_0t)\Bigg]} \Bigg] - 
\nonumber \\
&& 28  \Bigg[\frac{1}{1+\exp \Bigg[ \Bigg( \frac{1}{3}\frac{6^{1/2}35^{1/4}(\alpha_2^3 \beta^3 \gamma^2)^{1/4}}{\alpha_2 \beta} \Bigg)(x-\alpha_0t)\Bigg]} \Bigg]^2  + \nonumber \\
&& 28  \Bigg[\frac{1}{1+\exp \Bigg[ \Bigg( \frac{1}{3}\frac{6^{1/2}35^{1/4}(\alpha_2^3 \beta^3 \gamma^2)^{1/4}}{\alpha_2 \beta} \Bigg)(x-\alpha_0t)\Bigg]} \Bigg]^3 -
\nonumber \\
&& 24  \Bigg[\frac{1}{1+\exp \Bigg[ \Bigg( \frac{1}{3}\frac{6^{1/2}35^{1/4}(\alpha_2^3 \beta^3 \gamma^2)^{1/4}}{\alpha_2 \beta} \Bigg)(x-\alpha_0t)\Bigg]} \Bigg]^4 \Bigg \}
\end{eqnarray}
\end{description}
\section{Concluding remarks}
We discuss the  SEsM methodology  for obtaining exact solutions of nonlinear partial differential equations. The  goal of the old versions of this methodology was to help us to obtain particular exact 
solutions of nonlinear
nonitegrable partial differential equations of interest for mathematics, natural and social sciences.
We show that by appropriate extension of the methodology it can also lead to multisoliton solutions
of integravle differential equations.  We are sure that the methodology will be useful for the researchers who want to have a simple methodology for obtaining exact solutions of nonlinear partial differential equations.


\begin{thebibliography}{99}
\bibitem{a1}
M. Ausloos. Physica A \textbf{285}, 48 -- 65 (2000).
\bibitem{cs1}
L. Cameron, D. Larsen-Freeman. International Journal of Applied Linguistics \textbf{17}, 226 -- 239 (2007). 
\bibitem{cs2}
R. M. May, S. A. Levin, G. Sugihara. Nature \textbf{451},  893 -- 895 (2008).
\bibitem{vya1}
N. K. Vitanov,  E. D. Yankulova.  Chaos, Solitons \& Fractals \textbf{28}  768 -- 775 (2006).
\bibitem{a2}
K Ivanova, M. Ausloos. Physica A \textbf{274}, 349 -- 354 (1999).
\bibitem{vb1}
N. K. Vitanov, F. H. Busse.  Zeitschrift für Angewandte Mathematik und Physik ZAMP \textbf{48}, 310 -- 324 (1997).
\bibitem{vb1}
R. Borisov, N. K. Vitanov. AIP Conference Proceedings \textbf{2075}, 150001 (2019).
\bibitem{vb2}
N. K. Vitanov, R. Borisov. Journal of Theoretical and Applied Mechanics \textbf{48}, 74 -- 84 (2018).
\bibitem{cs3}
B. Brehmer. Acta Psychologica \textbf{81}, 211 -- 241 (1992).
\bibitem{va15}
N. K. Vitanov,  M. Ausloos.  Journal of Applied Statistics \textbf{42}, 2686 -- 2693 (2015).
\bibitem{a3}
R. Lambiotte, M. Ausloos. Journal of Statistical Mechanics: Theory and Experiment P08026 (2007).
\bibitem{bo}
B. Wang, Z. I. Dimitrova, N. K. Vitanov Journal of Theoretical and Applied Mechanics \textbf{49}, 136 -- 148 (2019).
\bibitem{knvit}
N. K. Vitanov. \emph{Science Dynamics and Research Production. Indicators, Indexes, Statistical Laws and Mathematical Models} (Springer, Cham, 2016).
\bibitem{vvx3}
N. K. Vitanov, R. Borissov. arXiv preprint arXiv:1803.05398 (2018).
\bibitem{cs4}
M. Mitchell. Artificial Intelligence \textbf{170}, 1194 -- 1212 (2006).
\bibitem{mars1}
M. Ausloos, A. Gadomski,  N. K. Vitanov.  Physica Scripta \textbf{89}, 108002 92014).
\bibitem{psv}
S. Panchev, T. Spassova, and N. K. Vitanov. Chaos, Solitons \& Fractals \textbf{33} 1658 -- 1671 (2007).
\bibitem{sak}
K. Sakai, S. Managi, N. K. Vitanov, K. Demura.  Nonlinear Dynamics, Psychology, and Life Sciences \textbf{11}, 253 -- 265 (2007).
\bibitem{cs5}
R. Mahon, P. McConney, R. N.Roy. Marine Policy \textbf{32},  104 -- 112 (2008).
\bibitem{a4x}
N. K. Vitanov, M. Ausloos, G. Rotundo. Advances in Complex Systems \textbf{15}, Supp. 01, 1250049 (2012)
\bibitem{vv16}
N. K. Vitanov, K. N. Vitanov.  Mathematical Social Sciences \textbf{80}, 108 -- 114 (2016).
\bibitem{vvv1}
N. K. Vitanov, K. N. Vitanov.  Physica A \textbf{509}, 635 -- 650 (2018).
\bibitem{vvv2}
N. K. Vitanov, K. N. Vitanov.  Physica A \textbf{490}, 1277 -- 1294 (2018).
\bibitem{vvx1}
N. K. Vitanov, K. N.  Vitanov. arXiv preprint arXiv:1807.08778 (2018).
\bibitem{vvx2} 
N. K. Vitanov, R. Borisov. arXiv preprint arXiv:1806.06659 (2018).
\bibitem{a4}
N. K. Vitanov,  K. N. Vitanov.  Physica A \textbf{527} 121174 (2019).

\bibitem{n1}
P. G. Drazin. \emph{Nonlinear Systems} (Cambridge University Press, Cambridge, UK, 1992).
\bibitem{n1x}
D. A. Hall. Journal of Materials Science \textbf{36},  4575 -- 4601 (2001).
\bibitem{n2}
A. J. Majda, R. V. Abramov, M. J. Grote. \emph{Information Theory and Stochastics for Multiscale Nonlinear Systems} (AMS, US, 2005).
\bibitem{n3}
I. Jordanov, E. Nikolova. Journal of Theoretical and Applied Mechanics \textbf{43}, No. 2, 69 -- 76 (2013). 
\bibitem{n4}
A. S. Pikovsky, D. L. Shepelyansky. Phys. Rev. Lett. \textbf{100}, 094101  (2008).
\bibitem{n5}
Z. I.  Dimitrova. Journal of Theoretical and Applied Mechanics \textbf{45}, No. 4, 79 -- 92 (2015).
\bibitem{psv}
S. Panchev, T. Spassova, and N. K. Vitanov. Chaos, Solitons \& Fractals \textbf{33} 1658 -- 1671 (2007).
\bibitem{sak}
K. Sakai, S. Managi, N. K. Vitanov, K. Demura.  Nonlinear Dynamics, Psychology, and Life Sciences \textbf{11}, 253 -- 265 (2007).
\bibitem{nikolova1}
E. V. Nikolova, I. P. Jordanov, Z. I. Dimitrova, N. K. Vitanov.   pp. 131 - 144 in K. Georgiev, 
M. Todorov,I. Georgiev (Eds.) Advanced Computing in Industrial Mathematics.Springer, Cham, 2018
\bibitem{n7}
Y. Niu,  S. Gong. Phys. Rev. A \textbf{73}, 053811 (2006).
\bibitem{vh99}
N. P. Hoffmann,  N. K. Vitanov. Physics Letters A \textbf{255}, 277 -- 286 (1999)
\bibitem{bo}
W. Bo, Z. I. Dimitrova, N. K. Vitanov. ArXiv:1906.00168 (2019).
\bibitem{cha2}
N. K. Vitanov,  K. N. Vitanov.  arXiv preprint arXiv:1904.11973 (2019).
\bibitem{n6}
Z. Dimitrova. Journal of Theoretical and Applied Mechanics \textbf{42}, No. 3, 3 -- 22 (2012).
\bibitem{vx1}
N. K. Vitanov.  Physics Letters A \textbf{248}  338 -- 346 (1998).
\bibitem{vx2}
T. Boeck, N. K. Vitanov.  Physical Review E \textbf{65}, 037203 (2002).
\bibitem{n8x}
E. Nikolova, E. Goranova, Z. Dimitrova. Comptes rendus de l'Academie Bulgare des sciences \textbf{69}, No. 9,
1213 -- 1222 (2016).
\bibitem{rb1}
R. Borisov, N. K. Vitanov. AIP Conference Proceedings \textbf{2075}, 150001 (2019).
\bibitem{rb2}
R. Borisov, N. K. Vitanov. arXiv preprint arXiv:1901.02361 (2019).
\bibitem{react}
N. K. Vitanov, K. N. Vitanov, Z. I. Dimitrova. arXiv preprint arXiv:1906.04828 (2019).
\bibitem{dush}
I. N. Dushkov, I. P. Jordanov. N. K. Vitanov. Mathematical Methods in the
Applied Sciences \textbf{41}, 8377 -- 8384 (2018)
\bibitem{n8}
A. Samoilenko, R. Petryshyn. \emph{Multifrequency Oscillations of Nonlinear Systems} (Kluwer, New York, 2004).
\bibitem{t1}
G. H. Golub, J. M. Ortega. \emph{Scientific Computing and Differential Equations} (Academic Press, San Diego, 1992).
\bibitem{t1x}
H. Kantz,  T. Schreiber. \emph{Nonlinear Time Series Analysis} (Cambridge University Press, Cambridge, UK, 2004).
\bibitem{t2}
P. J. Brockwell, R. A. Davis, M. V. Calder. \emph{Introduction to Time Series and Forecasting}. (Springer, New York, 2002).
\bibitem{t4}
N. K. Vitanov, M. Ausloos. Knowledge epidemics and populationdynamics models for describing idea diffusion, in \emph{Models of Science Dynamics}, edited by A.Scharnhorst, K Boerner, P. van den Besselaar,
(Springer, Berin, 2012) p.p. 65 -- 129.
\bibitem{jordi}
I. P. Jordanov, N. K. Vitanov. arXiv preprint arXiv:1808.02398 (2018)
\bibitem{t5}
F. Verhulst. \emph{Nonlinear Differential Equations and Dynamical Systems} (Springer, Berlin, 2006).
\bibitem{tse1}
N. K. Vitanov, Z. I. Dimitrova, T. I. Ivanova. ArXic 1708.01901 (2017).
\bibitem{el1} 
E. V. Nikolova. I. P. Jordanov, Z. I. Dimitrova, N. K. Vitanov.
arXiv preprint arXiv:1703.06429 (2017).
\bibitem{t6}
H. Kantz, D. Holstein, M. Ragwitz, N. K. Vitanov. Physica A \textbf{342},  315 -- 321 (2004).
\bibitem{el2} 
E.  Nikolova. I. P. Jordanov, N. K. Vitanov.
arXiv preprint arXiv:1701.02371 (2017).
\bibitem{t3}
K. T. Ashenfelter, S. M. Boker, J. R Waddell, N. Vitanov. Journal of Experimental Psychology: Human Perception and Performance \textbf{35},  1072 -- 1091 (2009).
\bibitem{gog1}
Z. I. Dimitrova. ArXiv 1509.08600 (2015).
\bibitem{gog2}
Z. I. Dimitrova. ArXiv 1303.0122 (2013).
\bibitem{t7}
M. H. Ernst. Physics Reports \textbf{78},  1 -- 171 (1981).
\bibitem{dv04}  
Z. I. Dimitrova, N. K. Vitanov. Theoretical Population Biology {\bf 66}, 1 -- 12 (2004).
\bibitem{t8}
T. B. Benjamin, J. L. Bona, J. J. Mahony, J. J. (1972).  Philosophical Transactions of the Royal Society of London. Series A, \textbf{272} 47 -- 78 (1972).
\bibitem{v00}
N. K. Vitanov. Physica D {\bf 136},  322 -- 339 (2000) 
\bibitem{el4}
E. V. Nikolova, D. Z. Serbezov, I. P. Jordanov. AIP Conference Proceedings
\textbf{2075}, 150003 (2019).
\bibitem{dv01}
Z. I. Dimitrova, N. K. Vitanov. Physica A  {\bf 300}, 91 -- 115 (2001) 
\bibitem{vda10}
N. K. Vitanov, Z. I. Dimitrova, M. Ausloos. Physica A, {\bf 389}, 4970 -- 4980 (2010).
\bibitem{t9}
D. B. Taulbee. Physics of Fluids A \textbf{4} 2555 -- 2561 (1992).
\bibitem{vhc}
N. K. Vitanov, A. Chabchoub,  N. P. Hoffmann. Journal of Theoretical and Applied Mechanics \textbf{43}, No.2, 43 -- 54 (2013).
\bibitem{dv00}
Z. I. Dimitrova, N. K. Vitanov. Phys. Lett A {\bf 272}, 368 -- 380 (2000)
\bibitem{t10a}
S. Grossberg. Neural Networks \textbf{1}, 17-61 (1981).
\bibitem{t10}
V. I. Arnol'd. \emph{Ordinary Differential Equations} (Spronger, Berlin, 1992).
\bibitem{hopf}
E. Hopf. Communications on Pure and Applied Mathematics, \textbf{3},  201 -- 230 (1950).
\bibitem{cole}
J. D. Cole. Quarterly of Applied Mathematics \textbf{9},   225 -- 236 (1951).	
\bibitem{ablowitz}
M. J. Ablowitz, D. J. Kaup, A. C. Newell, H. Segur. Studies in Applied  Mathematics,  \textbf{53},  
249 -- 315 (1974) .	
\bibitem{ac}
M. J. Ablowitz,  P. A. Clarkson. \emph{Solitons, Nonlinear Evolution Equations and Inverse Scattering}. (Cambridge University Press, Cambridge, UK, 1991).
\bibitem{gardner}
C. S. Gardner,  J. M. Greene, M. D. Kruskal, R. R. Miura.  Phys. Rev. Lett. \textbf{19}, 
1095 -- 1097 (1967).
\bibitem{hirota}
R. Hirota.   Phys. Rev. Lett. \textbf{27},   1192 -- 1194 (1971).
\bibitem{hirota1}
R. Hirota. \emph{The Direct Method in Soliton Theory}. (Cambridge University Press, Cambridge, UK, 2004).
\bibitem{tabor}
M. Tabor. \emph{Chaos and Integrability in Dynamical Systems}  (Wiley, New York, 1989).
\bibitem{ct1}
F. Carrielo, M. Tabor. Physica D,  \textbf{39},  77 -- 94 (1989).
\bibitem{ct2}
F. Carrielo, M. Tabor.  Physica D, \textbf{53}, 59 -- 70 (1991).
\bibitem{wtk}
J. Weiss, M. Tabor, G. Carnevalle. Journal of Mathematical Physics, \textbf{24},  522 -- 526 (1983).
\bibitem{k3}
N. A. Kudryashov. Physics Letters A, \textbf{155}, 269 -- 275 (1991).
\bibitem{k05}
N. A. Kudryshov. Chaos, Solitons \& Fractals \textbf{24},   1217 -- 1231 (2005).
\bibitem{kl08}
N. A. Kudryashov, N. B. Loguinova.  Applied Mathematics and Computation \textbf{205}, 361 -- 365 (2008).
\bibitem{k5a}
N. A. Kudryashov.  Physics Letters A, \textbf{342}, 99 -- 106 (2005).
\bibitem{k9}
N. A. Kudryashov, M. V. Demina. Applied Mathematics and Computation, \textbf{201},   551 -- 557 (2009).
\bibitem{k12a}
N. A. Kudryashov. Communications in Nonlinear Science and Numerical Simulation, \textbf{17}, 
26 -- 34 (2012).
\bibitem{k12}
N. A. Kudryashov. Communications in Nonlinear Science and Numerical Simulation, \textbf{17}, 
2248 -- 2253 (2012).
\bibitem{mv1}
N. Martinov, N. Vitanov. Journal of Physics A: Mathematical and General \textbf{25},  L51 -- L56 (1992).
\bibitem{mv2}
N. Martinov, N. Vitanov. Journal of Physics A: Mathematical and General \textbf{25}, L419 -- L426 (1992).
\bibitem{mv3}
N. K. Martinov, N. K. Vitanov. Journal of Physics A: Mathematical and General \textbf{27}, 4611 -- 4618 (1994).
\bibitem{mv4}
N. K. Martinov, N. K. Vitanov. Canadian Journal of Physics, \textbf{72}, 618 -- 624 (1994).
\bibitem{v98}
N. K. Vitanov. Proc. Roy. Soc. London A, {\bf 454},  2409 -- 2423 (1998)
\bibitem{mv5}
N. K. Vitanov. Journal of Physics A: Mathematical and General, \textbf{29},  5195 -- 5207 (1996).
\bibitem{1}
N. K. Vitanov, I. P. Joranov, Z. I. Dimitrova.  Communications in Nonlinear Science and Numerical Simulation 
\textbf{14}, 2379 -- 2388 (2009).
\bibitem{2}
N. K. Vitanov, I. P. Jordanov, Z. I. Dimitrova.   Applied Mathematics and Computation \textbf{215}, 2950-- 2964 (2009).
\bibitem{v10}
N. K. Vitanov. Communications in Nonlinear Science and Numerical Simulation
{\bf 15}, 2050 -- 2060 (2010).
\bibitem{vd10}
N. K. Vitanov,  Z. I. Dimitrova. Communications in Nonlinear Science and Numerical Simulation {\bf 15}, 2836 -- 2845  (2010).
\bibitem{vdk}
N. K. Vitanov, Z. I. Dimitrova, H. Kantz. Applied Mathematics and Computation, \textbf{216},  
2587 -- 2595 (2010).
\bibitem{v11}
N. K. Vitanov. Communications in Nonlinear Science and Numerical Simulation, \textbf{16}, 1176 -- 1185 (2011).
\bibitem{v11a}
N. K. Vitanov, Z. I. Dimitrova, K. N. Vitanov. Communications in Nonlinear Science and Numerical Simulation, \textsc{16},  3033 -- 3044 (2011).
\bibitem{v11b}
N. K. Vitanov.  Communications in Nonlinear Science and Numerical Simulation, \textsc{16},  4215 -- 4231 (2011).
\bibitem{pliska1}
N. K. Vitanov. Pliska Studia Mathematica Bulgarica \textbf{21}, 257 -- 266 (2012).
\bibitem{vdk13}
N. K. Vitanov, Z. I. Dimitrova, H. Kantz. Applied Mathematics and Computation, \textbf{219},  7480 -- 7492 (2013).
\bibitem{vdv13}
N. K. Vitanov, Z. I. Dimitrova, K. N. Vitanov.  Computers \& Mathematics with Applications, \textbf{66}, 1666 -- 1684 (2013).
\bibitem{vd14}
N. K. Vitanov, Z. I. Dimitrova. Applied Mathematics and Computation, \textbf{247},    213 -- 217 (2014).
\bibitem{vdv17}
N. K. Vitanov, Z. I. Dimitrova, T. I. Ivanova.  Applied Mathematics and Computation, \textbf{315},  372 -- 380 (2017).
\bibitem{vdv17p}
N. K. Vitanov, Z. I. Dimitrova, T. I. Ivanova. arXiv preprint arXiv:1708.01901 (2017).
\bibitem{vdv15}
N. K. Vitanov, Z. I. Dimitrova, K. N. Vitanov.  Applied Mathematics and Computation, \textbf{269},  363 -- 378 (2015).
\bibitem{vd18}
N. K. Vitanov,Z. I. Dimitrova. Journal of Theoretical and Applied Mechanics, Sofia, \textbf{48}, No. 1, 59  -- 68 (2018).
\bibitem{v19}
N. K. Vitanov. Pliska Studia Mathematica \textbf{30}, 29 -- 42 (2019). 	
\bibitem{v19a}
N. K. Vitanov. Journal of Theoretical and Applied Mechanics \textbf{49}, 107 - 122 (2019).
\bibitem{n17}
E. V. Nikolova, I. P. Jordanov, Z. I. Dimitrova, N. K. Vitanov.
AIP Conference Proceedings, \textbf{1895},  070002 (2017).
\bibitem{sesm1}
N. K. Vitanov. arXiv preprint arXiv:1908.07459 (2019).
\bibitem{sesm2}
N. K. Vitanov. arXiv preprint arXiv:1908.01075 (2019).
\bibitem{sesm3}
N. K. Vitanov. arXiv preprint arXiv:1906.08053  (2019)
\bibitem{sesm4}
N. K. Vitanov. arXiv preprint arXiv:1904.03481 (2019).
\bibitem{iv19}
I. P. Jordanov, N. K. Vitanov. Studies in Computational Intelligence \textbf{793}, 199-210 (2019).

\end{thebibliography}
\end{document}